\documentclass[%
 reprint, 
 superscriptaddress,
 amsmath,amssymb,
 aps,
floatfix, 
longbibliography,prb
]{revtex4-1}
\usepackage{graphicx}
\usepackage{bm}

\usepackage{xcolor}
\usepackage[colorlinks=true,citecolor=blue,linkcolor=magenta]{hyperref}

\usepackage[latin9]{inputenc}
\usepackage{amsmath}
\usepackage{amssymb}
\usepackage{graphicx}
\usepackage{verbatim}
\usepackage{bm}
\usepackage{color}

\usepackage{graphicx,epsfig,amsfonts,amssymb}
\usepackage{bm}% bold math
\usepackage{times}
\usepackage{lipsum}
\usepackage{verbatim}

\newcommand{\be}{\begin{eqnarray}}
\newcommand{\ee}{\end{eqnarray}} 

\newcommand{\bs}{\begin{equation}\begin{split}}
\newcommand{\es}{\end{split}\end{equation}}

\date{\today}

\begin{document}

\title{Asymmetry Amplification by a Nonadiabatic Passage through a Critical Point}
\author{Bhavay Tyagi}
\affiliation{Theoretical Division, Los Alamos National Laboratory, Los Alamos, New Mexico 87545, USA}
\affiliation{Department of Physics, University of Houston, Houston, Texas, 77204, USA}
\author{Fumika Suzuki}
\affiliation{Theoretical Division, Los Alamos National Laboratory, Los Alamos, New Mexico 87545, USA}
\affiliation{Center for Nonlinear Studies, Los Alamos National Laboratory, Los Alamos, New Mexico 87545, USA}
\author{Vladimir A. Chernyak}
\affiliation{Department of Chemistry and Department of Mathematics, Wayne State University, Detroit, Michigan, 48202, USA}
\author{Nikolai A. Sinitsyn}
\affiliation{Theoretical Division, Los Alamos National Laboratory, Los Alamos, New Mexico 87545, USA} 
\begin{abstract}
\noindent
We propose and solve a minimal model of dynamic passage through a  second-order phase transition in the presence of symmetry breaking interactions and no dissipation. Our model generalizes the Hamiltonian dynamics of the Painlev\'e-2 equation to the case with many degrees of freedom, while maintaining the integrability property.
The evolution eventually leads to a highly asymmetric state, no matter how weak the symmetry breaking parameter of the Hamiltonian is. This suggests a potential mechanism for strong asymmetry in the production of quasi-particles with nearly identical characteristics.  The model's integrability also yields exact exponents for the scaling of the density of the nonadiabatically  excited quasi-particles.
  \end{abstract}
\maketitle
\noindent
\section{Introduction}
 Only a small number of collective  variables matter when approaching a critical point of a phase transition. Therefore,    entire classes of systems at criticality are represented by basic field-theory models. 
 
 The simplest such models describe phase transitions in well-mixed dynamics, which is often found in computations by quantum annealing,  where interactions are represented by a dense graph of coupled qubits \cite{qarev,kamenevqa}. Near the critical points, the field theoretical approach  then has no spatial dimensions due to the long-range interactions in the computational Hamiltonian. The transitions through the critical points lead to nonadiabatic excitations that spoil the accuracy of computations, stimulating the interest in quantifying the corresponding nonadiabatic effects.

The most basic scalar field theory relevant to quantum  phase transitions (at zero temperature and without dissipation) has the  potential energy 
\begin{equation}
V({\bm \phi})= \frac{k|{\bm \phi}|^2}{2}+\frac{g|{\bm \phi}|^4}{2} , 
\label{phi40}
\end{equation}
where ${\bm \phi}=\phi_1+i\phi_2$ is a complex scalar field, and $\phi_{1,2}$ are  independent real fields. 
Here, we assume no spatial structure, meaning the system is well mixed.

Allow the parameter $k$ in~(\ref{phi40}) to vary with time. For $k\ge 0$, $V({\bm \phi})$ has one unique minimum, and for $k<0$ there is a continuum of minima, namely,
\begin{eqnarray}
  \label{phi40-0}
 \phi_{\rm min}&=& 0, \quad {\rm for } \quad k\ge 0, \\
   \label{phi40-1}
    \phi_{\rm min}&=&\sqrt{|k|/(2g)}e^{i\varphi}, \quad {\rm for } \quad k<0,
\end{eqnarray}
where $\varphi \in [0,2\pi)$ is an arbitrary phase.

%The dynamic passage through a critical point is encountered not only in quantum annealing but also in cosmological waterfall models, e.g.,  of the inflationary epoch of our universe~\cite{linde,tahion,Reheating}, in which case the time-dependence of $k(t)$ follows from the coupling of the phenomenological ``waterfall" field ${\bm \phi}$ with the time-dependent inflaton field. The phase transition then leads to the condensation of ${\bm \phi}$ that terminates inflation. 

In experiments with ultra-cold atoms, an analogous phase transition is encountered in coherent reactions between atomic and molecular condensates, whose dynamics near the Feshbach resonance is described by a second-order phase transition with the energy functional  (\ref{phi40}), where ${\bm \phi}$ would be an amplitude of either molecular or atomic condensates~\cite{gurarielz,itin}. 
In both quantum annealing and stimulated reactions in ultra-cold atoms, the passage through the critical point is not adiabatic, so the system does not have to follow the minimum of energy~(\ref{phi40}) during its time evolution. However, far from the critical point, the adiabaticity conditions for quantum coherent evolution are restored. Therefore, to describe the nonadiabatic effects, it is sufficient to consider a linearized time-dependence of $k(t)$ near the critical point: 
\begin{equation}
k(t) \approx-\beta t,
\label{linear}
\end{equation}
where $\beta \equiv -k'(0)>0$ is the characteristic rate of the transition through the critical point that is reached at time $t=0$. This approximation is analogous to the Landau-Zener approximation near an avoided crossing of quantum energy levels.

Assuming that ${\bm \phi}$ is dimensionless, e.g., representing the amplitude of a condensate, the model~(\ref{phi40}) with (\ref{linear}) has an intrinsic time scale, $\tau=g/\beta$, during which the critical effects are essential, around  $t=0$. One famous prediction, which is universal for all 2nd order phase transitions,  is the power law scaling for the number of nonadiabatic excitations that are induced by the passage through the critical point: $n_{ex}\propto\tau^{-\nu}$, where $\nu$ is the  Kibble-Zurek exponent~\cite{kibble,zurek}.

 In this article, we  provide a detailed description of the nonadiabatic passage through the phase transition in the field theory with the energy potential $V({\bm \phi})$ and an additional term that  breaks the  symmetry between $\phi_1$ and $\phi_2$ by introducing a  mass difference between them: 
\begin{equation}
 V({\bm \phi})=-\frac{\beta t |{\bm \phi}|^2}{2} +\frac{g |{\bm \phi}|^4}{2} +\varepsilon\left( \varepsilon_1 \phi_1^2+\varepsilon_2 \phi_2^2 \right), \quad \varepsilon_1<\varepsilon_2,
    \label{V-broken}
\end{equation}
 where $\varepsilon_{1,2} \sim O(1)$ are dimensionless parameters, and $\varepsilon$ is the characteristic energy scale for the symmetry breaking. The evolution starts as $t\rightarrow -\infty$ near the global  ground state at $\phi_1=\phi
_2=0$, and proceeds to $t\rightarrow +\infty$, while the entire system remains isolated from environment.

 \section{Saddle-point equations}

%e.g., in models of matter-antimatter asymmetry induced by a weakly broken CP-symmetry~\cite{CP1}.

Let ${\bm \pi}(t)$ be the canonically conjugate momentum variable to ${\bm \phi}(t)$. Adding a dynamic contribution 
to the effective action $S[{\bm \pi}(t) {\bm \phi}(t)]$, the model acquires the form of the $0+1$-dimensional quantum field theory 
\begin{equation}
S[{\bm \pi}(t) {\bm \phi}(t)]= \int  \left[ \sum_{k=1,2} \pi_k \, d\phi_k -H({\bm \pi},{\bm \phi}) \, dt \right],
\label{action}
\end{equation}
where the Hamiltonian $H({\bm \pi},{\bm \phi})$ has a  kinetic energy  that depends on the momenta $\pi_k$:
$$
H({\bm \pi},{\bm \phi})= \frac{\pi_1^2+\pi_2^2}{2m} +V({\bm \phi}),
$$
where $m$ is a new  parameter, and $V({\bm \phi})$ is given by Eq.~(\ref{V-broken}).
The saddle-point equations of motion are given by~\cite{LL} 
\begin{equation}
\frac{d\phi_k}{dt} = \frac{\partial H}{\partial \pi_k}, \quad \frac{d\pi_k}{dt} = -\frac{\partial H}{\partial \phi_k}.
\label{canon1}
\end{equation}

In what follows, we will show that the equations of motion (\ref{canon1}) can be solved analytically without further approximations, i.e., for arbitrary parameter values $g$, $\beta$, $m$, $\varepsilon$, and $\varepsilon_{1,2}$. This solution provides an opportunity to answer certain fundamental questions about the  nonadiabatic dissipationless phase transitions. First, it is {\it a priori} not known whether such dissipationless dynamics lead to the creation of a condensate near the minimum of $V({\bm \phi})$ because the energy here is not conserved. In fact, any discrete quantum system placed in linearly time-dependent fields remains near the initial state in the limit of a fast transition through all resonances. Generally, this leads to a highly excited final state, e.g., as in the Landau-Zener process \cite{Damsky2005}, or in the quantum Ising spin chain with a transverse field that sweeps very quickly through the critical point \cite{Dzyarmaga2005,sinitsyn21prl}.
Does this property apply to the dissipationless classical system with action (\ref{action}) in the strongly nonadiabatic regime, $\beta^{1/2}\gg g,\varepsilon$? We will show that Eqs.~(\ref{canon1}) always lead to the creation of the condensate as $t\rightarrow +\infty$. 

Second, an interesting regime corresponds to very weak asymmetry   $\varepsilon$ in comparison to the other energy scales: 
\begin{equation} 
\begin{split}
    \varepsilon &\ll \beta^{1/2}\text{ and }\varepsilon \ll g.  
\end{split}
\label{asymm}
\end{equation}
Even though the condensate is created, the strong nonadiabatic effects may, in principle, produce the final state with the condensate field ${\bm \phi}$ having a time-dependent phase $\varphi$, which 
would reduce the effect of the asymmetry parameter to a small modulation of the condensate phase dynamics. This behavior is indeed observed in our system at small $\varepsilon$ during a long time interval $\propto 1/\sqrt{\varepsilon}$ after the passage through the critical point at $t=0$. 
However, our main result is that the saddle-point equations for the time-evolution with the potential energy (\ref{V-broken}) always lead to a highly asymmetric final state as $t\rightarrow +\infty$, regardless of how small $\varepsilon$ is. This occurs because the solution at $\varepsilon=0$ is unstable. For arbitrarily small but nonzero $\varepsilon$, the asymmetry effects accumulate over time and eventually the dynamics is directed toward the state with a strongly broken symmetry between $\phi_1$ and $\phi_2$.

Finally, the solution of our model enables us to characterize the nonadiabatic excitations analytically exactly. Thus, the model~(\ref{phi40})  has two types of excitations over the energy minimum~(\ref{phi40-1}) after the phase transition: the massless Goldstone and the massive Higgs bosons. One might expect that the nearly-massless Goldstone bosons are easier to produce but we will demonstrate that the Higgs bosons are produced in much larger quantities.

%, as depicted for a specific trajectory in Fig.~\ref{fig:1}(a,b)
\section{Asymptotically adiabatic dynamics}
Let us rescale variables by factors, $\lambda$, $u$ and $v$, so that 
$$ 
t\rightarrow \lambda t, \quad \pi_k=v P_k, \quad \phi_k = u X_k, \quad k=1,2. 
$$
In terms of $P_k$ and $X_k$, Eqs.~(\ref{canon1}) are still canonical with a new Hamiltonian ${\cal H}({\bf X}, {\bf P})=\lambda H(u{\bf X}, v{\bf P}) /uv$, where ${\bf X}=(X_1,X_2)$ and ${\bf P}=(P_1,P_2)$. We choose the factors so that the symmetric terms of the Hamiltonian have only numerical coefficients:
\begin{equation}
{\cal H}=\frac{{\bm P}^2}{2} -t \frac{{\bm X}^2}{2} +\frac{{\bm X}^4}{2} + \varepsilon \left(e_1 X_1^2+e_2 X_2^2 \right),
\label{hamu}
\end{equation}
which is achieved at $\lambda=(m/\beta)^{1/3},\,u=m^{1/6}\beta^{1/3}/\sqrt{g},\,v=m^{5/6}\beta^{2/3}/\sqrt{g}$, so that $e_k=\varepsilon_k\lambda u/v = \varepsilon_k/ (\beta^2 m)^{1/3}$.

%%%%%%%%%%%%%%%%%%%%%%%%%%%%%%%%%%%%%
\begin{figure}
 \includegraphics[width=0.99\columnwidth]{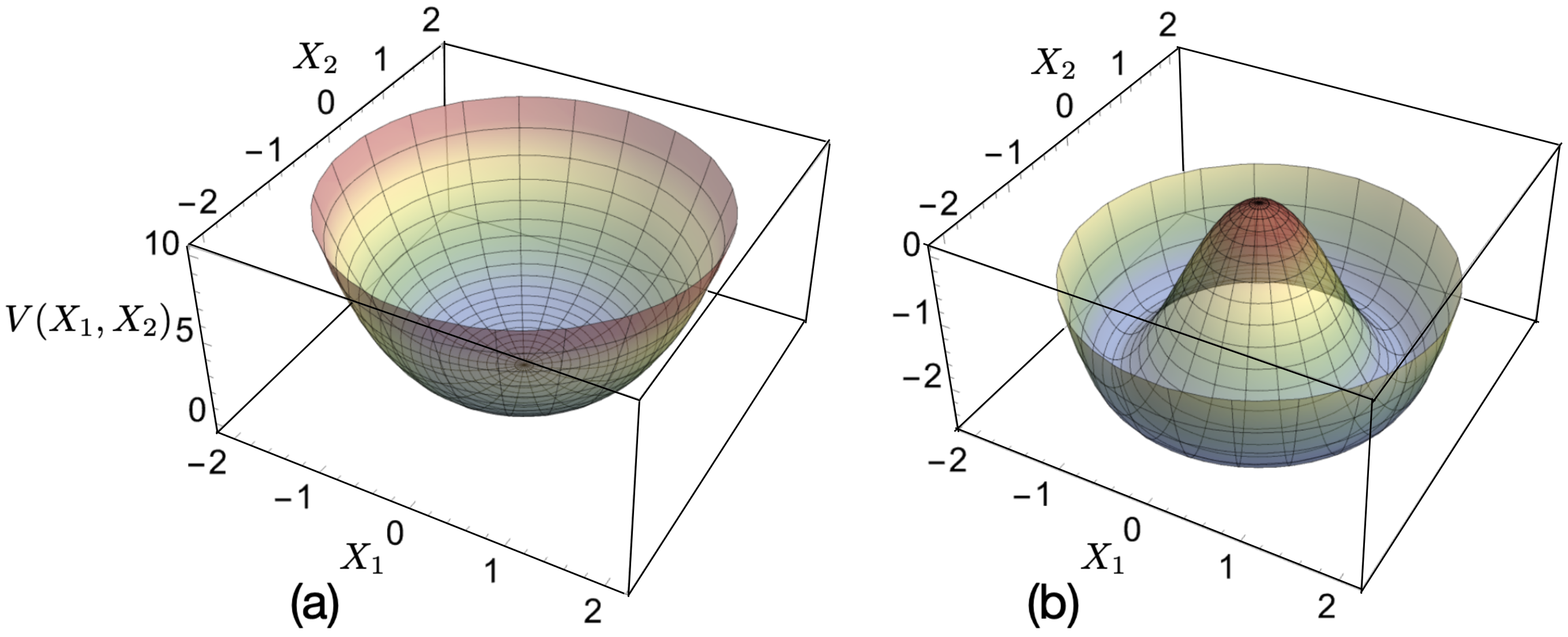}
    \caption{The potential energy  
 $V(X_1,X_2)=-t\frac{|{\bf X}|^2}{2}+\frac{{\bf X}^4}{2}$, where ${\bf X}^2\equiv X_1^2+X_2^2$, and 
    (a) $t=-1<0$, (b)  $t=4.5>0$. Adding small nonzero $\varepsilon$  leads only to a small distortion of $V(X_1,X_2)$. }
    \label{energy-landscape}
\end{figure}
%%%%%%%%%%%%%%%%%%%%%%%%%%%%%%%%%%%%%

Figure~\ref{energy-landscape} demonstrates the potential energy landscape for negative and positive $t$. 
As $t\rightarrow \pm \infty$, the potential energy along the radial direction near the minimum resembles that of a harmonic oscillator with growing frequency $\omega \propto \sqrt{|t|}$. Hence,  the evolution is asymptotically adiabatic, thereby conserving the adiabatic invariants~\cite{LL}:  
$$
I_k=\frac{1}{\pi}\int_{x_k^{-}}^{x_{k}^+} dX_k \, P(X_k), \quad k=1,2, 
$$
where $x_{k}^{\pm}$ are the turning points of the classical trajectory~(\ref{canon1}). In quantum mechanics, this invariant is related to the number of excitations $n_k$ of the quantum fields $\phi_k$ over the ground level. Thus, in terms of the physical variables the adiabatic invariant is given by 
$$
{\cal I}_k\equiv \frac{1}{\pi} \int_{\phi_k^{-}}^{\phi_k^+} \pi_k \, d\phi_k=\hbar (n_{k}+1/2). 
$$
We assume that the system is in the vacuum state as $t\rightarrow -\infty$, which corresponds to ${\cal I}_{1,2}^{-\infty}  = \hbar/2$. The corresponding canonically conjugate angle variables $\varphi_{1,2}$  are not determined initially. The following, both analytical and numerical, results describe the averaging over uniformly distributed $\varphi_{1,2} \in [0,2\pi)$.
Our change of variables was noncanonical, so
\begin{equation}
{\cal I}_k^{\pm \infty}  = I_k^{\pm \infty} uv = m\beta I_k^{\pm \infty}/g,
\label{convert}
\end{equation}
and the evolution with ${\cal H}$ starts with a small, $O(\hbar)$, value $I_k^{-\infty}=\hbar g/(2\beta m) \ll 1$.

To characterize the nonadiabatic effects we  must find the asymptotic value $\langle I_k\rangle$ averaged over the initial conditions as $t\rightarrow +\infty$: $\langle I_k^{+\infty} \rangle =I_k^{ad} +\Delta I_k$, where  $I_k^{ad}$ is the adiabatic invariant that would be found in the limit of the adiabatic evolution.
The number of nonadiabatic excitations of $\phi_k$ is then given by 
\begin{equation}
n_k=\beta m \Delta I_k/(\hbar g).
\label{ex-num}
\end{equation}

Thus, our model resembles a scattering problem, in which the dynamics is asymptotically trivial. In particular, it conserves the number of excitations defined by Eq.~(\ref{ex-num}) as $t\rightarrow \pm \infty$. All nontrivial effects are localized near the time moment of the passage through the critical point at $t=0$.

\section{Integrability conditions}
Our most nontrivial finding is that the function
\begin{eqnarray} 
\label{hcom2} 
 &&{\cal H}' \equiv 
t(e_1X_1^2+e_2X_2^2)-2\varepsilon(e_1^2X_1^2+e_2^2X_2^2)-
\\\nonumber&&e_1(P_1^2+X_1^4+X_1^2X_2^2)-e_2(P_2^2+X_1^2X_2^2+X_2^4)+\frac{L^2}{2\varepsilon },
\end{eqnarray}
where $L=P_2X_1-P_1X_2$ is the angular momentum, satisfies the following relations with ${\cal H}$, from Eq~(\ref{hamu}):
\begin{eqnarray}
\label{int-conds}
 \nonumber \frac{\partial {\cal H}}{\partial \varepsilon} &=&\frac{\partial {\cal H}'}{\partial t}, \\
  \{{\cal H},{\cal H}' \}&\equiv& \sum_{k=1,2} \frac{\partial {\cal H}}{\partial X_k} \frac{\partial {\cal H'}}{\partial P_k}-\frac{\partial {\cal H}}{\partial P_k} \frac{\partial {\cal H'}}{\partial X_k}=0. 
\end{eqnarray}

%%%%%%%%%%%%%%%%%%%%%%%%%%%%%%%%%%%%%
\begin{figure}
 \includegraphics[width=0.7\columnwidth]{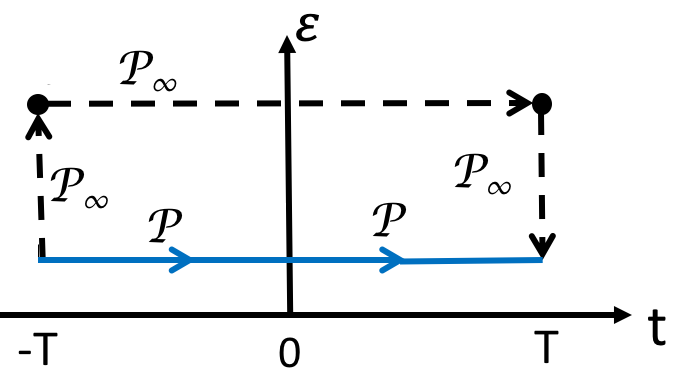}
    \caption{The physical path, ${\cal P}$,  of the evolution in the $(t,\varepsilon)$ plane goes through a critical point at $t\approx 0$ at fixed $\varepsilon\ll \beta/g$. Here, $T\rightarrow \infty$.
The path ${\cal P}_{\infty}$ initially lifts $\varepsilon$ with the Hamiltonian ${\cal H'}$ to a large value $\varepsilon \gg \beta/g$ at formally infinite $|t|$. The horizontal part of ${\cal P}_{\infty}$ produces the evolution with ${\cal H}$ at large rather than small fixed $\varepsilon$. The second vertical leg of ${\cal P}_{\infty}$ brings the system back to small $\varepsilon$ at infinite $t$.}
    \label{fig:2}
\end{figure}
%%%%%%%%%%%%%%%%%%%%%%%%%%%%%%%%%%%%%
\noindent 
This makes the 1-form 
$$
\Omega={\cal H} \,dt+{\cal H}'\,d\varepsilon
$$
closed. In Appendix~\ref{IC-section}, we show that the integration of the classical equations of motion can then be extended to a two-time $\tau$-plane $(\tau_1,\tau_2)\equiv (t,\varepsilon)$ with a two-Hamiltonian $({\cal H}_1,{\cal H}_2)=({\cal H},{\cal H'})$, so that
$$
\frac{\partial X_k}{\partial \tau_j}=\frac{\partial {\cal H}_j}{\partial P_k}, \quad \frac{\partial P_k}{\partial \tau_j}=-\frac{\partial {\cal H}_j}{\partial X_k}, \quad k,j=1,2.
$$
\noindent
Analogously to the integrable multistate Landau-Zener models \cite{commute,parallel}, the result of the evolution along an arbitrary path in the $\tau$-plane then depends solely on the initial and final two-time-points, and does not depend on the choice of the path connecting them, as long as this path does not cross any singular point of ${\cal H}'$ at $\varepsilon=0$. Our original physical evolution corresponds to the path ${\cal P}$, shown in Fig.~\ref{fig:2}.  To avoid any complex dynamics encountered along ${\cal P}$ near $t=0$, we deform it into the path ${\cal P}_{\infty}$, which bypasses this region from a large distance.

Along the left vertical leg of ${\cal P}_{\infty}$, the dynamics is dominated by the $t$-dependent terms because $t\rightarrow -\infty$. Keeping only the most relevant terms, we find for this piece
$$
\frac{d^2X_k}{d\varepsilon^2} +(2 e_k)^2|t|X_k=0, \quad k=1,2, 
$$
which corresponds to the harmonic oscillators with  formally infinite frequencies.

Thus, the first vertical leg produces only the adiabatic dynamics that conserves $I_{1,2}$. It merely brings the system to the point with large $\varepsilon$, such that $\varepsilon e_k \gg \beta/g$. Analogously, along the other vertical part of ${\cal P}_{\infty}$, the nonadiabatic transitions are suppressed by a formally infinite value of $t$.

\begin{figure*}
\includegraphics[width=2\columnwidth]{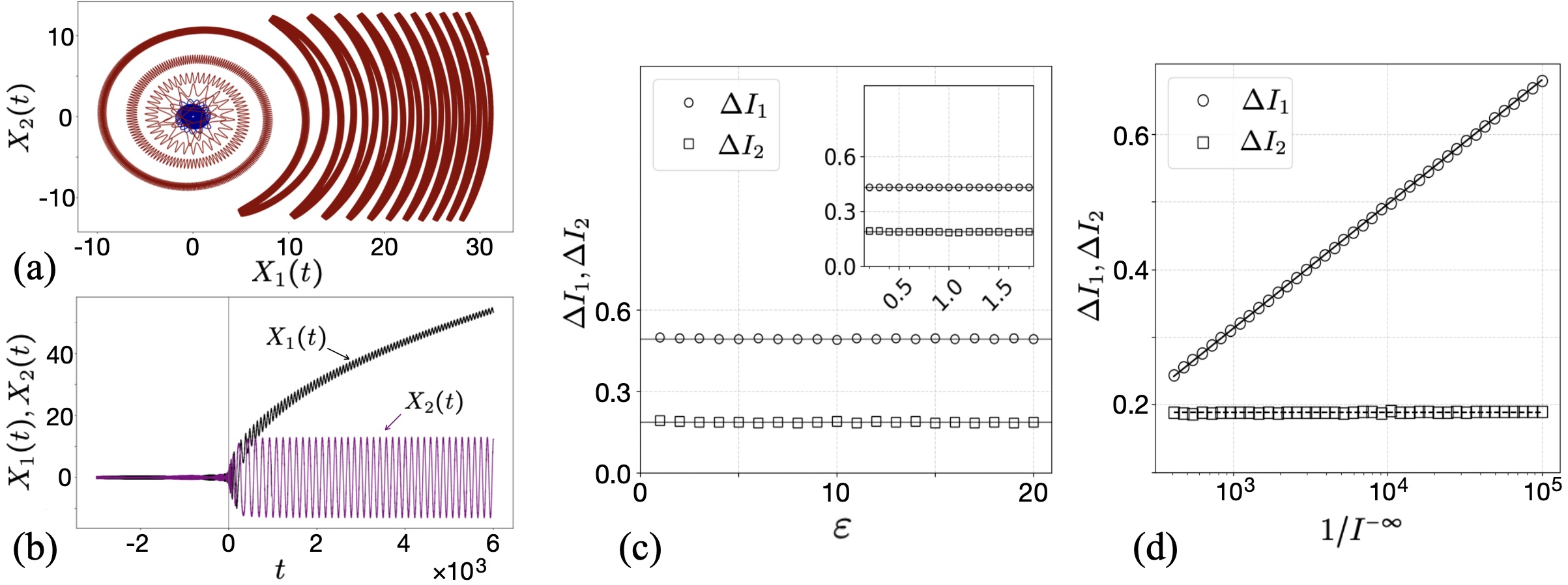}
 \caption{
    (a) Trajectory $\{X_1(t),X_2(t)\}$ for $t\in [-3000,2000]$ (blue: $t<0$, red: $t>0$) and the Hamiltonian (\ref{hamu});  $\varepsilon=0.001$, $e_1=-1/2$, $e_2=1/2$, $X_1(t=-3000)=X_2(t=-3000)=-0.1$, $P_1(t=-3000)=-20$, $P_2(t=-3000)=20$. (b) Coordinates $X_1(t)$ (black) and $X_2(t)$ (purple) as functions of time for the same $\varepsilon$,~$e_{1,2}$, and initial conditions. After the phase transition at $t=0$, both $X_1$ and $X_2$ pick up the amplitude with initially equal rate but for longer times, $t>1/\sqrt{\varepsilon}$, $X_1(t)$ keeps growing, while $X_2(t)$ oscillates near $\langle X_2 \rangle =0$. 
    (c) Numerical check for $\varepsilon$-independent behaviour of $\Delta I_{1,2}$ at $I^{-\infty}_{1,2}=0.0001$. The inset resolves the interval $\varepsilon \in [0.2,1.8]$. The time interval for simulations is $t\in (-500,3000)$ and the time-step is $dt=0.0005$. 
    Each point is the average of the final adiabatic invariant over $6000$ trajectories with different initial angles, in the action-angle variables, taken from a uniform distribution $\varphi_{1,2} \in (0,2\pi)$. (d) Numerical confirmation for Eq.~(\ref{corrected-actions}): the logarithmic dependence of $\Delta I_1$ on $I_{1,2}^{-\infty}= I^{-\infty}$ and a constant value for   $\Delta I_2$. Here, $\varepsilon=1$ and the other parameters are as in (c). 
 %%%%%%%%%%%%%%%%%%%%%%%%%%%%%%%%%%%
  }
    \label{fig:1} 
\end{figure*}

The evolution along the horizontal part of ${\cal P}_{\infty}$ proceeds during $t\in (-\infty, \infty)$ with the Hamiltonian ${\cal H}$ at very large $\varepsilon$. The main consequence of the model's integrability is then the independence of the evolution result, for the Hamiltonian~(\ref{hamu}), of the strength of the symmetry breaking parameter, $\varepsilon>0$, which we verified numerically in Fig.~\ref{fig:1}(c). Hence, it is sufficient for our goals to study the case of $\varepsilon \gg g,\beta^{1/2}$ to describe arbitrary symmetry breaking. %%%%%%%%%%%%%%%%%%%%%%%%%%%%%%%%%%%%
\section{Main results}
At large $\varepsilon$, $X_1$ enters the phase transition at $t_1=2\varepsilon e_1$, which is long before the critical moment for $X_2$, at $t_2=2\varepsilon e_2$. Near $t=t_1$, the variable $X_2$  experiences a strong confinement due to the quadratic potential $\sim \varepsilon (\varepsilon_2-\varepsilon_1)X_2^2$, and for large $\varepsilon$ performs fast oscillations, whose effect on $X_1$ averages to zero.  The evolution of $X_1$ is then governed by the truncated to $(X_1,P_1)$ Hamiltonian ${\cal H}_1=P_1^2/2-sX_1^2/2 +X_1^4/2$, where $s=t-2\varepsilon e_1$. The corresponding equation of motion
\begin{equation}
 d^2X_1/ds^2 =s X_1-2X_1^3
    \label{P2-1}
\end{equation}
 is the Painlev\'e-2 equation, whose solution has known asymptotic behavior \cite{P2-book} as $s\rightarrow +\infty$:
\begin{equation}
\begin{split}
    X_1(s) &=\pm \sqrt{\frac{s}{2}}\\
    &\pm \rho (2s)^{-1/4} 
    \cos \left(\frac{2\sqrt{2}}{3}s^{3/2}-\frac{3}{2}\rho^2 \ln s +\theta(\varphi_0) \right),
\end{split} 
\label{plus-inf}  
\end{equation}
where $\rho=\sqrt{2I^{+\infty}}$, and the term $\pm \sqrt{s/2}$ in Eq.~(\ref{plus-inf}) corresponds to the time-dependent positions of the new potential minima. The  term $\pm \sqrt{\frac{s}{2}}$, signifies that the variable $X_1$ is growing in absolute value with time, which corresponds to creation of the condensate of the field $\phi_1$ as $t\rightarrow +\infty$. This behavior is found for arbitrary values of the parameters in the original model. Thus, the behavior of quantum systems in the fast transition limit is not reproduced by the saddle-point equations. A full quantum mechanical treatment of our field theory model would be needed to describe the extremely nonadiabatic quantum regime, in which discreetness of the spectrum would prevent the creation of the condensate.

In Appendix~\ref{Asol-sec}, we derive the change of the adiabatic invariant for Eq.~(\ref{P2-1})
at small initial $I_1^{-\infty}$:
 \begin{equation}
\Delta I_1 = I_1^{+\infty}-I_1^{-\infty}/2=\frac{1}{4\pi}\ln \left(\frac{1}{2\pi I_1^{-\infty}}\right).
\label{deltaI-1}
\end{equation}
After the transition through the first critical point, at $t=2\varepsilon e_1$, along the horizontal leg of ${\cal P}_{\infty}$, the variable $X_1$ is dominated by the growing contribution $X_1^2\sim s/2=(t-2\varepsilon e_1)/2$. Substituting this into the Hamiltonian ${\cal H}$, we find the effective Hamiltonian for $X_2$:
$$
{\cal H}_2=\frac{P_2^2}{2} +\varepsilon (e_2-e_1) X_2^2+X_2^4/2.
$$
\noindent
Importantly, this Hamiltonian does not explicitly depend on $t$. Since $e_2>e_1$, small excitations are confined by a nearly-harmonic potential with the frequency
\begin{equation}
\omega_2=\sqrt{2\varepsilon (e_2-e_1)}.
\label{om-x2} 
\end{equation}
In the rest of ${\cal P}_{\infty}$, the  evolution for $X_2$ does not go through a new critical point.  As $t\rightarrow +\infty$, the variable $X_2$ oscillates near the local minimum, as shown for the numerically obtained trajectory in Fig.~\ref{fig:1}(a,b).

Since the variable $X_2$ is not growing permanently with time, {\it   the field $\phi_2$, asymptotically as $t\rightarrow +\infty$,  does not form a condensate}. No matter how small the physical parameter $\varepsilon$ may be in comparison to the interaction and dynamic parameters, the macroscopic condensate is always created solely by the field $\phi_1$. The inverse of the frequency in Eq.~(\ref{om-x2}) corresponds to the time-scale of this asymmetry accumulation. Hence, to observe this effect, the time-dependence in the mass term should be $k \sim \beta/\omega$.

The nonadiabatic excitations, described by  $\Delta I_2$,  are small but finite. They follow from the fact that, after the symmetry breaking for $X_1$, the frequency of the oscillatory component of $X_1$ in (\ref{plus-inf}) grows as $\omega_1 \approx \sqrt{2(t-2\varepsilon e_1)}$. The leading nonlinear coupling is a $1{:}2$ resonance, $\omega_1=2\omega_2$, which occurs at the distant moment $t_{\rm res}=2\varepsilon(2e_2-e_1)$. At this resonance, the nonadiabatic excitations for $X_1$ and $X_2$ can exchange their powers due to the nonlinear interaction terms. 
We found numerically (Fig.~\ref{fig:1}(d)) that for small initial $I_2^{-\infty}=I_1^{-\infty}$, this resonance produces only subdominant excitations that appear as a non-logarithmic correction, so that 
\begin{equation}
 \Delta I_1=\frac{1}{4\pi} \left[\ln \left(\frac{1}{2\pi I_1^{-\infty}}\right)-c_1\right], \quad \Delta I_2=\frac{c_2}{2\pi},
 \label{corrected-actions}
\end{equation}
where we determined  that $c_1=1.14$ and $c_2=1.19$ are constants that do not depend on the initial conditions as long as $I_1^{-\infty}=I_2^{-\infty} \ll 1$. By restoring the physical variables via (\ref{convert}) and (\ref{ex-num}), we find the  excitation numbers 
\begin{equation}
  n_1=\frac{\beta m}{4\pi g\hbar} \left[\ln \left(\frac{\beta m}{\pi \hbar g}\right)-c_1\right], \quad n_2=\frac{c_2\beta m}{2\pi g \hbar}.
\label{corrected-numbers}
\end{equation} 
\noindent
Since $\phi_1$ forms the condensate, its oscillatory excitations are the Higgs bosons. Their average number is given by $n_1$ in (\ref{corrected-numbers}). 
The excitations of the field $\phi_2$ are massless in the limit $\varepsilon \rightarrow 0$, so they are the Goldstone bosons. 

For specific applications, the parameters $m$ and $g$ must be determined by further details of the models, beyond the effective field theory, and the constants $c_{1,2}$ can be affected by the quantum corrections to our assumption that initially ${\cal I}_{1,2}^{-\infty}=\hbar/2$. However, the scaling, $n_1\sim \Gamma \ln \Gamma$  and $n_2\sim \Gamma \equiv \beta m/(g\hbar)$, is our prediction for the Kibble-Zurek exponent of the universality class described by the mean field theory~(\ref{hamu}).
 
 \section{Discussion}
Our scaling predictions can be verified through computations by quantum annealing  and experiments
with ultracold atoms and molecular condensates ~\cite{itin,kamenevqa}. One analytical result, which has been extensively tested by quantum annealing experiments, is the prediction of $\nu=1/2$  Kibble-Zurek scaling exponent for the number of elementary excitations in 1-dimensional Ising model in a transverse field \cite{Damsky2005,Dzyarmaga2005}. This exponent is anomalous. Its value reflects the integrable and 1-dimensional character of the Ising chain.
In practice, the most interesting computational annealing problems  correspond to dense graphs of inter-qubit interactions far from the integrability. 
We showed here that the most elementary, and therefore the most fundamental, well mixed field theory predicts the Kibble-Zurek exponent $\nu\sim 1$ for such systems.  

Other applications of our result may emerge in the problems  of separating or producing particles with very similar characteristics, such as 
 the problem of separation of isotopes with similar masses \cite{sinitsyn21prl}. Our results also lead to speculations about the origin of the matter-antimatter asymmetry in the universe, given the smallness of the CP-violation in the Standard Model. We have shown that an elementary field theory model can explain a strong asymmetry in particle production from a vacuum that becomes unstable after the phase transition. 

Finally, our approach demonstrates that the integrability of the Painlev\'e equations, whose solutions are among the most important special functions of mathematical physics, can be generalized to multi-dimensional Hamiltonian equations, with the possibility of connecting asymptotic solutions as $t\rightarrow \pm \infty$. Our model~(\ref{hamu}) can be generalized to a  system with more than two degrees of freedom:
\begin{equation}
{\cal H}=\sum_{k=1}^N \left[ \frac{ P_k^2}{2} -t \frac{ X_k^2}{2} + \varepsilon e_k X_k^2 \right] +\frac{\left(\sum_{k=1}^N X_k^2 \right)^2}{2},
\label{hamu2}
\end{equation}
where $N$ is an arbitrary integer number, $t$ is time and $e_1<e_2<\ldots<e_N$ are arbitrary parameters. Another Hamiltonian
\begin{eqnarray}
\label{hcomN}
\nonumber {\cal H}' \!\equiv \!\sum_{k=1}^{N} \left[
-2\varepsilon e_k^2X_k^2+
e_k(tX_k^2 \!-\!P_k^2-X_k^2\sum_{j=1}^N X_j^2) \right]\!+\!\frac{{\bf L}^2}{2\varepsilon },
\end{eqnarray}
where $L_{jk}=X_jP_k-X_kP_j$ are components of the antisymmetric angular-momentum tensor, and
$
{\bf L}^2\equiv \sum_{j,k=1}^{N}L_{jk}^2
=\sum_{1\le j<k\le N}(X_jP_k-X_kP_j)^2 
$. ${\cal H}' $ satisfies the integrability conditions with ${\cal H}$: $\{{\cal H},{\cal H}'\}=0$ and $\partial_{\varepsilon} {\cal H}=\partial_t {\cal H}'$.
It should be insightful in the future to find more integrable nontrivial pairs of Hamiltonians that depend only linearly or inversely linearly on $t$ and $\varepsilon$. Our theory states that under such conditions the asymptotic behavior of variables at $t=\pm \infty$ can be connected analytically. This should help us grasp the  dynamic nonperturbative effects in more complex critical phenomena.

\appendix

%\newline
\begin{acknowledgments}

 This work was supported primarily by the U.S. Department of Energy, Office of Science, Office of Advanced Scientific Computing Research, through the Quantum Internet to Accelerate Scientific Discovery Program, and in part by the U.S. Department of Energy, Office of Science, Basic Energy Sciences, under Award Number DE-SC0022134. B.T. also acknowledges  support from NSF grant CHE-2404788. F.S. acknowledges support from the Los Alamos National Laboratory LDRD program under project number 20230049DR and the Center for Nonlinear Studies under project number 20220546CR-NLS.
\end{acknowledgments}

\appendix
\section{Integrability conditions}
\label{IC-section}
 Here, we review the notion of an integrable family of classical Hamiltonians. Let $X$ be a phase space, i.e., a smooth manifold, equipped with a Poisson bracket
\begin{eqnarray}
\label{Poisson-bracket} \{f, g\} = \sum_{ij} \omega_{ij} (x) \frac{\partial f}{\partial x_i} \frac{\partial g}{\partial x_j},
\end{eqnarray}
with $f, g \in {\cal H}$, with ${\cal H}$ being the vector space of smooth functions on $X$, and let $\{H_{\alpha}  (\bm{\tau})\, | \, \alpha = 1, \ldots, n, \, | \, \bm{\tau} \in U\}$ be a smooth family of classical Hamiltonians over an open single connected region $U \subset \mathbb{R}^n$. Then any smooth path $\bm{\tau} : [0, 1] \to U$ defines the time-dependent Hamiltonian $H (t) = \sum_{\alpha} \dot{\tau}_{\alpha} H_{\alpha} (\bm{\tau} (t))$. Our goal is to identify the explicit conditions for integrability, i.e., that the solution of the Hamilton equation
\begin{eqnarray}
\label{Hamilton-eq} \dot{x}_j = \sum_k \omega_{jk} (x(t)) \frac {\partial H (\bm{\tau} (t), x)}{\partial x_k},
\end{eqnarray}
depends on the initial condition, as well as the initial and final points of a path $\tau$, and is independent of a particular choice of a path. Since the Hamilton equation is equivalent to the Liouville equation
\begin{eqnarray}
\label{Liouville-eq} \dot{f} = \{H (\bm{\tau} (t)), f\}
\end{eqnarray}
for the evolutions of functions/distributions $f \in {\cal H}$, the integrability conditions can be alternatively identified for Eq.~(\ref{Liouville-eq}), rather than~(\ref{Hamilton-eq}). Indeed, the Poisson bracket equip ${\cal H}$ with a structure of an infinite-dimensional Lie algebra, and
\begin{eqnarray}
\label{Liouville-connection} \nabla_{\alpha} f = \frac{\partial f}{\partial \tau_{\alpha}} - \{H_{\alpha}, f\},
\end{eqnarray}
define a non-abelian connection/gauge field/covariant derivative in a trivial infinite-dimensional vector bundle ${\cal H} \times U \to U$ with the fiber ${\cal H}$ and structure/gauge group $G$ being the group of canonical diffeomorphisms/transformations of $X$. According to~\cite{commute}, the integrability conditions are given by vanishing of the curvature ${\cal F}$ of the connection, defined in Eq.~(\ref{Liouville-connection}):
\begin{eqnarray}
\label{Liouville-connection-curv}  {\cal F}_{\alpha\beta} = \frac{\partial H_{\alpha}}{\partial \tau_{\beta}} - \frac{\partial H_{\beta}}{\partial \tau_{\alpha}} - \{H_{\alpha}, H_{\beta}\} = 0.
\end{eqnarray}

%A derivation that does not involve infinite-dimensional spaces and thus can be much easier converted to a mathematically rigorous proof is represented in appendix~\ref{sec:diff-geometry-derive}

\section{Asymptotic solution of  Painlev\'e-2}
\label{Asol-sec}
\noindent
We will follow closely the analysis in \cite{itin}, which is based on the known asymptotic solutions of the Painlev\'e-2 \cite{P2-book}.
Starting as
$s\rightarrow - \infty$ the Painlev\'e-2 
equation (\ref{P2-1}) describes trivial oscillations in a harmonic potential: 
$$
X_1(s) =\alpha (-s)^{-1/4}\sin \left(\frac{2}{3}(-s)^{3/2}+\frac{3}{4} \alpha^2\ln (-s) +\varphi_0\right),
$$
For $s\rightarrow +\infty$: 
\begin{equation}
\begin{split}
    &X_1(s) =\pm \sqrt{\frac{s}{2}}\\
    &\pm \rho (2s)^{-1/4}
    \cos \left(\frac{2\sqrt{2}}{3}s^{3/2}-\frac{3}{2}\rho^2 \ln s +\theta(\varphi_0) \right),
\end{split}
\label{plus-inf-2}
\end{equation}
where the term $\pm \sqrt{s/2}$ corresponds to the time-dependent positions of the new potential minima, and 
\begin{equation}
\rho^2=\frac{1}{\pi} \ln \frac{1+|p|^2}{2|\Im (p)|},\quad  p=\sqrt{e^{\pi \alpha^2}-1}e^{i\theta'(\varphi_0)},
\label{rho-def}
\end{equation}
where $\theta$, $\theta'$ are phases whose precise values are not needed for our discussion because we should average the final result over the initial phase $\varphi_0$, and hence over $\theta$ and $\theta'$.

%The first contribution in Eq~(\ref{plus-inf}) shows that $X_1$ after the phase transition  condenses to the value $\pm\sqrt{s/2}$, that corresponds to the minimum (\ref{phi40-1}) at $\varphi=0$. 
The oscillatory terms in $X_1$ are due to the excitations. 
Long before the critical point the system is in a parabolic potential with frequency $\omega_{-\infty}=\sqrt{|s|}$, and long after the critical point, $\omega_{+\infty}=\sqrt{2s}$. For the harmonic oscillator, the half-axis of the oscillation and the adiabatic invariant, $I_1$, are related by $I_1=\Delta X_1^2\omega/2$ \cite{LL}, so we can now identify 
\begin{equation}
I_1^{-{\infty}}=\alpha^2/2, \quad I_1^{+\infty} = \rho^2/2.
\end{equation}
The nonadiabatic shift of this invariant is given by $\Delta I_1 = I_1^{+\infty}-I_1^{-\infty}/2$, where $1/2$ is a purely geometric effect that reflects the change of the phase space volume by a trajectory at crossing the separatrix of the double well potential. We then write $p=|p|e^{i\pi\xi}$, where $\xi$ is treated as a uniformly random number $\xi \in (0,1)$. Substituting this into Eqs.~(\ref{plus-inf-2})-(\ref{rho-def}), and averaging over  $\xi$, we find 
\begin{equation}
\Delta I_1=-\frac{1}{4\pi}\ln \left(1-e^{-2\pi I_1^{-{\infty}}}\right).
\end{equation}
For $I_1^{-\infty}\ll1$, we arrive at  
Eq.~(16) in the main text.


\begin{thebibliography}{100} 


\bibitem{qarev} 
A. Rajak, S. Suzuki, A. Dutta,  and B.~K. Chakrabarti. Phil.l Trans. Royal Soc. A: Math., Phys. and Eng. Sci., A {\bf 381}:20210417 (2022). {\it Quantum annealing: an overview}.

\bibitem{kamenevqa} R. Xie and A. Kamenev. Phys. Rev. A {\bf 110}, 032418 (2024). {\it Quantum Hopfield model with dilute memories}.


 
%\bibitem{linde}  Andrei Linde. Phys. Rev. D {\bf 49}, 748 (1994). {\it Hybrid inflation}.

%\bibitem{Reheating}  L. Kofman, A. Linde, and A. A. Starobinsky. Phys. Rev. Lett. {\bf 73}, 3195 (1994). {\it Reheating after Inflation}.

%\bibitem{tahion} E. J. Copeland, S. Pascoli, and A. Rajantie. Phys. Rev. D {\bf 65}, 103517 (2002) {\it Dynamics of tachyonic preheating after hybrid inflation}.

\bibitem{gurarielz} A.~Altland, V.~Gurarie, T.~Kriecherbauer, and A.~Polkovnikov.
Phys. Rev. A {\bf 79}, 042703 (2009). {\it Nonadiabaticity and large fluctuations in a many-particle Landau-Zener problem.}

\bibitem{itin} A.P. Itin, P. {T\"orm\"a}. Preprint arXiv:0901.4778 (2009) {\it  Dynamics of quantum phase transitions in Dicke and Lipkin-Meshkov-Glick models.}; V.~Ganesh Sadhasivam, F.~Suzuki, B.~Yan, N.~A.~Sinitsyn. Nature Comm. {\bf 15}, Article number: 10246 (2024). {\it Parametric tuning of dynamical phase transitions in ultracold reactions.}


\bibitem{kibble} T. W. B. Kibble  J. Phys. A: Math. Gen. {\bf 9} 1387 (1976). {\it Topology of cosmic domains and strings.}

\bibitem{zurek} W.~H. Zurek. Phys. Rep. {\bf 276}, 177 (1996). {\it Cosmological experiments in condensed matter systems}. 

\bibitem{Damsky2005} B. Damsky. Phys. Rev. Lett. {\bf 95}, 035701 (2005). The simplest quantum model supporting the Kibble-Zurek mechanism of topological defect production: Landau-Zener transitions from a new perspective.

\bibitem{Dzyarmaga2005} J. Dzyarmaga. Phys. Rev. Lett. {\bf 95}, 245701 (2005). Dynamics of quantum phase transition: exact solution in quantum Ising model.

\bibitem{sinitsyn21prl}
B. Yan, V. Y. Chernyak, W. H. Zurek, and N. A. Sinitsyn. 
Phys. Rev. Lett. {\bf 126}, 070602 (2021). {\it Nonadiabatic Phase Transition with Broken Chiral Symmetry.}

%\bibitem{CP1} V. M. G. Silveira, C. A. Z. Vasconcellos, E. G. S. Luna, D. Hadjimichef. J. High Energy Phys. {\bf 2021} 285 (2021). {\it Matter{-}antimatter asymmetry and non-inertial effects}.

\bibitem{LL} L. D. Landau and E. M. Lifshitz. Vol.~{\bf I}, Nauka, Moskva (1973). {\it Classicheskaya Mehanika.} (in Russian).

\bibitem{commute} N. A. Sinitsyn, E. A. Yuzbashyan, V. Y. Chernyak, A. Patra, and C. Sun.  Phys. Rev. Lett. {\bf 120}, 190402 (2018). {\it Integrable time-dependent quantum Hamiltonians.}

\bibitem{parallel} V.~Y.~Chernyak, F.~Li, C.~Sun, and N.~A.~Sinitsyn. J. Phys. A: Math. Theor. {\bf 53} 295201 (2020). {\it Integrable multistate {Landau-Zener} models with parallel energy levels.}

\bibitem{P2-book} 
A. S. Fokas, A. R. Its, A. A. Kapaev, and V. Yu. Novokshenov. {\it Painlev\'e Transcendents: The Riemann-Hilbert Approach}. Mathematical Surveys and Monographs (Mathematical Surveys and Monographs, 128), American Mathematical Society (October 10, 2006).




%\bibitem{suppl} In Supplemental Material, we follow Refs.~\cite{itin,commute,P2-book} to review classical integrability conditions and derive the change of the adiabatic invariant using known asymptotic solutions of the Painlev\'e-2 equation.

\end{thebibliography}
\end{document}